\documentclass[journal]{vgtc}                
\ifpdf
  \pdfoutput=1\relax                   
  \usepackage{graphicx}                
  \DeclareGraphicsExtensions{.pdf,.png,.jpg,.jpeg} 
\else
  \usepackage{graphicx}                
  \DeclareGraphicsExtensions{.eps}     
\fi%

\graphicspath{{figures/}{pictures/}{images/}{./}} 

\usepackage{amssymb}
\usepackage{booktabs}
\usepackage{microtype}                 
\PassOptionsToPackage{warn}{textcomp}  
\usepackage{textcomp}                  
\usepackage{mathptmx}                  
\usepackage{times}                     
\usepackage{cite}                      
\usepackage{tabu}                      
\usepackage{todonotes}
\usepackage{booktabs}                  
\usepackage[utf8]{inputenc}
\usepackage[T1]{fontenc}
\usepackage{stmaryrd}

\definecolor{green}{rgb}{0, 0.5, 0}

\onlineid{1109}

\vgtccategory{VAST papers}
\vgtcpapertype{System}

\title{II-20: Intelligent and pragmatic analytic categorization of image collections}

\author{Jan Zah\'{a}lka, Marcel Worring, \textit{Senior Member, IEEE}, and Jarke J. van Wijk}
\authorfooter{
\item
 Jan Zah\'{a}lka is with the Czech Technical University in Prague. E-mail: jan@zahalka.net.
\item
 Marcel Worring is with the University of Amsterdam. E-mail: m.worring@uva.nl.
\item
 Jarke J. van Wijk is with the Eindhoven University of Technology. E-mail: j.j.v.wijk@tue.nl.
}

\shortauthortitle{Zah\'{a}lka \MakeLowercase{\textit{et al.}}: II-20: Intelligent and pragmatic analytic categorization of image collections}

\abstract{In this paper, we introduce \textbf{II-20} (Image Insight 2020), a multimedia analytics approach for analytic categorization of image collections. Advanced visualizations for image collections exist, but they need tight integration with a machine model to support the task of analytic categorization. Directly employing computer vision and interactive learning techniques gravitates towards search. Analytic categorization, however, is not machine classification (the difference between the two is called the \emph{pragmatic gap}): a human adds/redefines/deletes categories of relevance on the fly to build insight, whereas the machine classifier is rigid and non-adaptive. Analytic categorization that truly brings the user to insight requires a flexible machine model that allows dynamic sliding on the exploration-search axis, as well as semantic interactions: a human thinks about image data mostly in semantic terms. II-20 brings three major contributions to multimedia analytics on image collections and towards closing the pragmatic gap. Firstly, a \textbf{new machine model} that closely follows the user’s interactions and dynamically models her categories of relevance. II-20’s machine model, in addition to matching and exceeding the state of the art’s ability to produce relevant suggestions, allows the user to dynamically slide on the exploration-search axis without any additional input from her side. Secondly, the dynamic, 1-image-at-a-time \textbf{Tetris metaphor} that synergizes with the model. It allows a well-trained model to analyze the collection by itself with minimal interaction from the user and complements the classic grid metaphor. Thirdly, the \textbf{fast-forward interaction}, allowing the user to harness the model to quickly expand (``fast-forward'') the categories of relevance, expands the multimedia analytics semantic interaction dictionary. Automated experiments show that II-20’s machine model outperforms the existing state of the art and also demonstrate the Tetris metaphor’s analytic quality. User studies further confirm that II-20 is an intuitive, efficient, and effective multimedia analytics tool.
} 


\CCScatlist{ 
 \CCScat{K.6.1}{Management of Computing and Information Systems}%
{Project and People Management}{Life Cycle};
 \CCScat{K.7.m}{The Computing Profession}{Miscellaneous}{Ethics}
}

\teaser{
  \centering
  \includegraphics[width=0.85\linewidth]{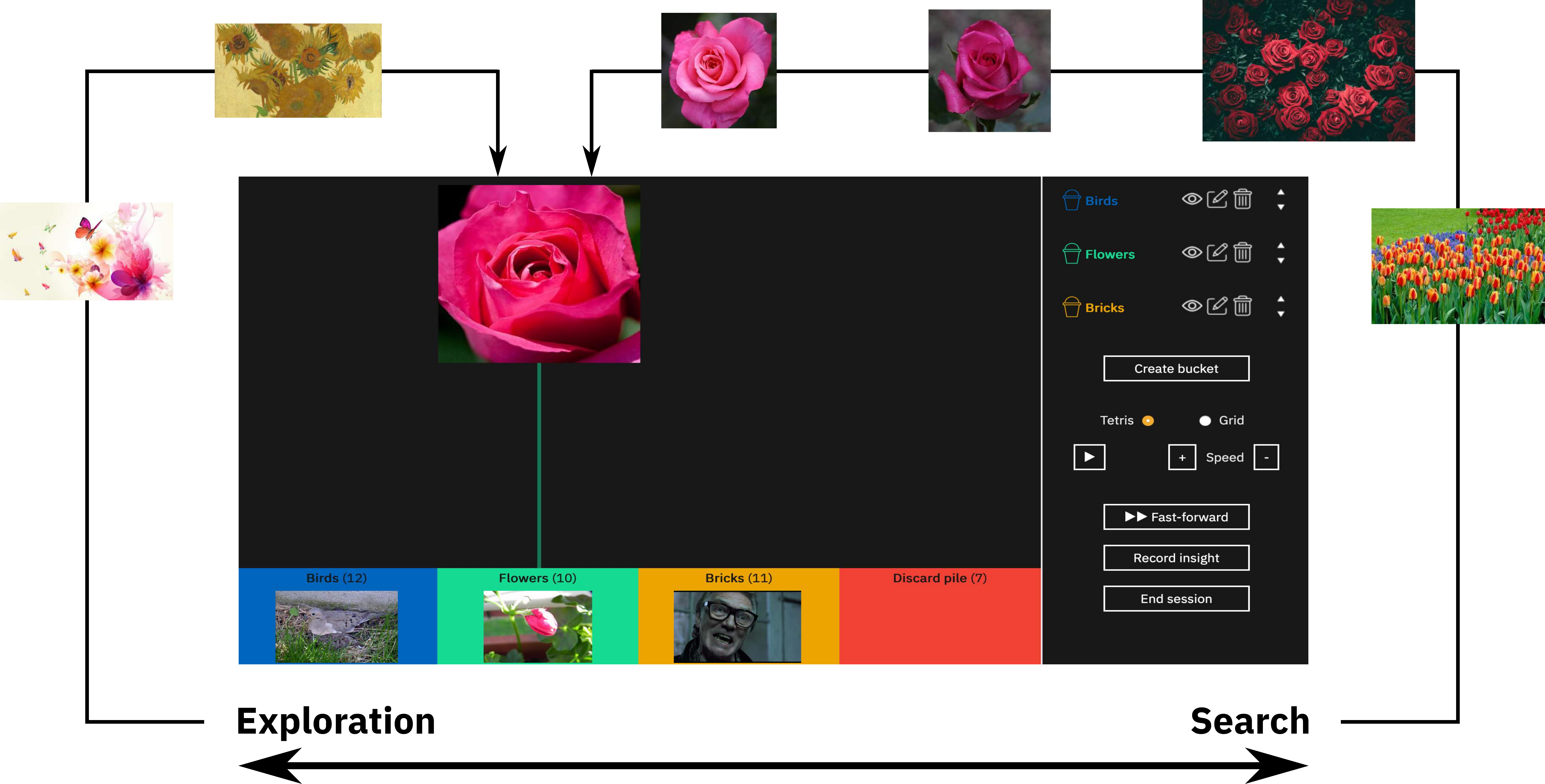}
  \caption{II-20's novel model fully supports flexible analytic categorization of image collections, closing the pragmatic gap. The user can categorize or discard images, and the system intelligently chooses the level of exploration and search for the categories. Pictured: the novel Tetris interface metaphor.}
  \label{fig:teaser}
}

\vgtcinsertpkg

\begin{document}


\firstsection{Introduction}

\maketitle

The growing wealth and importance of multimedia data (images, text, videos, audio, and associated metadata) is evident. Processing them meaningfully and efficiently has become crucial for an increasing number of domains, e.g., media and news, forensics, security, marketing, and health. The ubiquity and availability of cameras have made  casual content more important than ever. Social networks are a multi-billion dollar industry and user-contributed content is valuable. Visual data (images and videos) are at the core of the multimedia explosion and there is a great need for advanced analytics of visual data collections.

In recent years, our ability to automatically process large volumes of visual data has improved greatly. The chief reason is the dramatic increase of semantic quality of machine feature representations, spearheaded by deep neural networks \cite{krizhevsky12}. In many tasks, deep nets approach or surpass human capabilities, e.g., in object recognition (with equal train and test noise levels) \cite{dnns_better_than_humans_recog}. The semantic gap \cite{smeulders00} has been closed for many tasks and is rapidly closing for others. The quality and accessibility of advanced classifiers and indexes have entrenched the search engine as the golden standard for analyzing image collections.



However, not all multimedia analytics tasks boil down to just search. In a general analytics task on multimedia data, the user dynamically oscillates between exploration and search on the \emph{exploration-task axis} \cite{zahalka14}. Examples of tasks that are not purely search include:

\begin{itemize}
\item[T1)] \emph{Structuring the collection} --- make sense of what is in a collection with unknown contents, and structure it based on multiple categories of relevance.
\end{itemize}
Examples of this would be a marketing specialist analyzing her company brand’s perception on social media, or a quality control manager visually inspecting finished products for flaws.
\begin{itemize}
\item[T2)] \emph{Finding needles in the haystack} --- in a collection with only a small portion of relevant items, find them based on complex, often domain- and expertise-dependent semantics. 
\end{itemize}
For example, a forensics analyst trying to establish whether there is criminal content on a suspect’s seized computer.
\begin{itemize}
\item[T3)] \emph{Subjective/highly contextual content retrieval} --- labeling of the content into categories which can only be defined by the user as they are subjective or contextual. 
\end{itemize}
In this case as the notion of relevance cannot be defined beforehand or grounded objectively, the content-based indexes have trouble matching the user’s input to their dictionary. An example here is “show me art that I like” which does not match predefined content labels very well.


The need to support varied tasks can be addressed by employing the visual analytics approach, supporting knowledge/insight gain by tight integration of advanced visualizations with a machine model \cite{keim_va_mastering, keim08, keim_new}. Image collection analytics belongs to multimedia analytics \cite{chinchor10}, which has a number of specifics: among others, strong focus on semantics, high information bandwidth, and difficult summarization.

In general, multimedia analytics tasks can be modeled as \emph{analytic categorization}, in which the user defines the categories of relevance herself on the fly, and the model adapts to them as the session progresses \cite{zahalka14}. This is different from the classic machine learning classification, and the difference between the two is the \emph{pragmatic gap} \cite{zahalka14}. Analytic categorization requires support for multiple categories of relevance at once, creating/redefining/deleting categories on the fly during the analytic session, and strong emphasis on interactivity: the user interactions drive the categorization and vice versa, and they complete in interactive time. New visualizations and models built specifically around tight integration of the two and support of analytic categorization are needed.

There are approaches that incorporate interactive model building to cover a wider range of the exploration-search axis. To advance the analytic session, they usually make use of a rich set of filters on the data (e.g.,  \cite{iclic, pivot_tables, barthel_maps_graphs, photospread, urban_beautification, miranda20}), an interactive (multimodal) learning model (e.g., \cite{informedia, city_melange}), or a combination of both. Whilst these techniques go beyond search, on the exploration-search axis, they tend to lean towards search anyway: they simply fetch what the users are looking for or what they found relevant previously. To date, multimedia analytics retains an hourglass interface-model structure: a wide array of visualizations on the one hand, a wide array of automatic multimedia analysis techniques on the other, and a narrow set of interactions between the two --- filter, search, interactive learning (relevance feedback, active learning).

To enable meaningful interaction, semantic interactions are vital to multimedia analytics. Semantic interactions translate user interactions performed on high-level visual artifacts in the interface to low-level analytic model adjustments, coupling cognitive and computational processes \cite{sem_interactions}. The user does not train and adjust the model directly, but rather interacts within her domain of expertise, the model uses those interactions to improve. Developing new such interactions and thus widening the hourglass would improve multimedia analytics capabilities further.

To address the above challenges, we present \textbf{II-20} (Image Insight 2020), a multimedia analytics approach for image collections that brings the following contributions:

\begin{itemize}
\item A \textbf{new analytic categorization model} that supports multiple categories of relevance and dynamically slides on the exploration-search axis without explicit user involvement. The model is fully interactive even on large (> 1M) datasets and metaphor-agnostic. To the best of our knowledge, II-20's model is the first to \emph{fully} support analytic categorization of image collections.
\item The \textbf{Tetris metaphor} that streams the images in one-by-one, with the user steering them to the correct categories of relevance. As the model learns, it starts playing the Tetris game by itself with the user only correcting the model's mistakes. The metaphor is tightly integrated with the model with a clear benefit to the user: the number of interactions required from her is inversely proportional to the quality of the model whilst providing the same or better analytic outcome.
\item The \textbf{fast-forward interaction} that allows the user to swiftly categorize a large number of images at once based on the current state of the model.
\end{itemize}

The rest of the paper is organized as follows. \autoref{sec:rel_work} overviews the related work. \autoref{sec:ii20} describes II-20. \autoref{sec:exp_setup} outlines the experimental setup, results are discussed in \autoref{sec:discussion}. \autoref{sec:conclusion} concludes the paper.

\section{Related work}
\label{sec:rel_work}
Analytic categorization of image collections is iterative, requiring tight integration between the visual metaphor and the machine model that provides image suggestions. This is in line with the established visual analytics theory \cite{keim_va_mastering, keim_new, keim08, pirolli_card}. True support of analytic categorization as a task involves semantic interactions (this challenge is shared with general visual analytics \cite{sem_interactions}), dynamic sliding on the exploration-search axis, and closing the pragmatic gap \cite{zahalka14}. In this section, we review related work on the constituent parts of a multimedia analytics system (interface and model) and on means of integrating the two.

There is a great variety of visual metaphors available. The classic, time-tested approach used by the vast majority of systems visualizing image collections is the grid. Grids score near-perfectly on efficiency of screen space utilization and are very intuitive. They can be enhanced to convey collection structure  \cite{quadrianto10, barthel_maps_graphs, zavesky08}. Beyond grids, there are many other metaphors, such as spreadsheet-based that integrate the image content tightly with metadata \cite{pivot_tables, mediatable, photospread}, semantic-navigation-based that allow the user to pursue threads of interest, often semantic \cite{brivio_voronoi, threadeus}, or even metadata-driven \cite{iffi, semantic_vis}. Rapid serial visual presentation (RSVP) presents images dynamically, flashing images or video clips in a fast-paced manner, with the user providing simple, rapid response \cite{rsvp, rsvp_eindhoven, extreme_vr}. There are plenty of metaphors with various niches.

Models supporting multimedia analytics can be split roughly into index-based and interactive-learning-based (hybridization possible). Index-based techniques precompute a collection index which is used for filtering and/or search queries. The basic, yet still effective approach is the metadata-based index. Content-based indexing requires feature and/or concept label extraction, and the contemporary computer vision standard is to use deep convolutional neural networks \cite{krizhevsky12}. The features (esp. semantically meaningful ones, such as concept labels) can be used as metadata (e.g., \cite{iclic}) or to build a content-based index to fuel search. Indexing approaches include clustering-based approaches such as product quantization \cite{pq} or hash-based approaches \cite{hashing, lsh}. The current state of the art offers a broad range of techniques that establish a semantic structure of the collection. Relying on indexing alone in multimedia analytics, however, reduces analytics to just search. The model is rigid, non/adaptive, and does not address the pragmatic gap.

Interactive-learning-based approaches collect feedback from the user in the form of explicit ``relevant'' and ``not relevant'' labels, then train a new model based on those labels, rank the data according to the new model, and suggest more relevant items. Each interaction round should happen in interactive time. Following visual analytics theory \cite{interaction_defined}, this means operating in the real-time ($<$0.1 s latency) or direct manipulation (0.1 -- 2-3 s) regime. There are two dominant approaches. Firstly, relevance feedback \cite{rf_survey}, which suggests images the model deems most relevant. Secondly, active learning \cite{settles_al, aggarwal_al}, which suggests images the model is least sure about. This maximizes the model's learning gain and minimizes the number of user interactions. Algorithmically, most of the techniques come from the 2000s (the aforementioned surveys provide a good overview). In the 2010s, interactive learning struggled with the rapid increase in data scale. Recently, it has been improved to work on modern large-scale collections by introducing efficient compression \cite{blackthorn} and clustering \cite{exquisitor}. Interactive learning is adaptive, dynamic, and flexible: it learns only from the user, making it a good fit for closing the pragmatic gap. On its own, however, it still gravitates towards search, and is limited by latency: there is only so much that can be computed in interactive time.

In the 2000s and 2010s, there have been a number of systems that integrate advanced visualizations with machine learning-based models in both visual analytics \cite{sota_va_ml} and multimedia analytics \cite{zahalka14}. Moreover, visual analytics has been employed to \emph{explain} machine learning models. A recent notable instance is the effort to explain deep neural nets \cite{dnn_va}. In most of the visual analytics systems, users directly manipulate the machine learning model, which is useful for the data scientist, but might be difficult for an analyst who is not a machine learning expert. The multimedia analytics systems, where semantic navigation is of paramount importance, usually operate with a narrow semantic interaction dictionary: `filter'', ``search'', and ``perform interactive learning''. Additional semantic interactions would definitely be a big boon for both visual and multimedia analytics \cite{sem_interactions, zahalka14}.

As discussed in the introduction, II-20 brings three main contributions. II-20's machine model combines the advantages of index-based and interactive-learning-based approaches. By flexibly supporting dynamic sliding on the exploration-search axis, it is to the best of our knowledge the first model closing the pragmatic gap \cite{zahalka14}. The Tetris metaphor, beyond expanding the family of metaphors for image collections, has a tighter integration with the model than others, decreasing the number of interactions as the model improves. Finally, the fast-forward interaction expands the semantic interaction dictionary, answering a clear visual and multimedia analytics research challenge \cite{sem_interactions, zahalka14}.
\section{II-20}
\label{sec:ii20}
\begin{figure}
    \centering
    \includegraphics[width=\columnwidth]{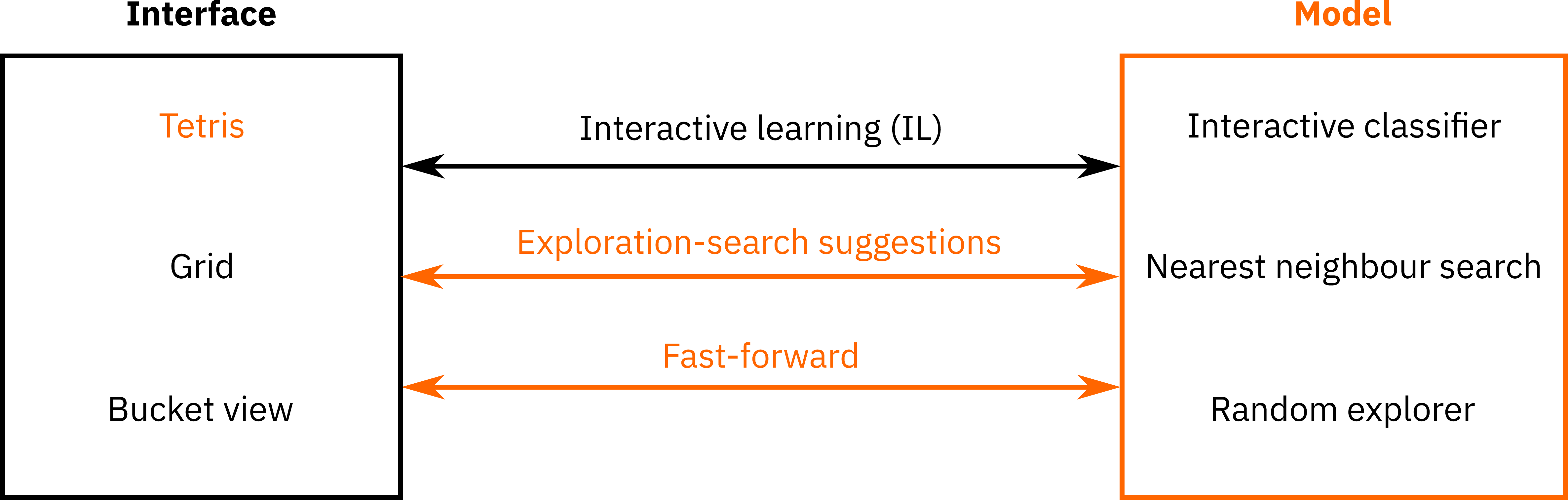}
    \caption{II-20's interface-model scheme. The components innovated by II-20 are coloured orange.}
    \label{fig:ii20_scheme}
\end{figure}

II-20 is tailored for full support of analytic categorization, defined as the task of assigning images $i_1, ..., i_n$ from the collection $I$ into analytic categories, which we henceforth call \emph{buckets} consistently with the terminology introduced in related work \cite{mediatable, pivot_tables}. The machine model requires the images to be represented with a semantic feature representation. 

The support for flexible buckets is formalized as follows. Let $B$ denote the set of user-defined buckets and $b \in B$ an individual bucket. To cater for the pragmatic gap, $B$ is a mutable set: the user can create, redefine, activate/deactivate, and remove buckets at any time throughout the analytics session. In II-20, the user can have between 1 and 7 buckets active at any given time, which is consistent with related work on visualization theory \cite{7_categories}. Individual buckets are mutable as well, images can be added, removed, and transferred between buckets at any time. In addition to $B$, II-20 adds the implicit \emph{discard pile} bucket ($d$) which at all times contains images that were discarded by the user (marked not relevant). The user can add images to $d$ and restore them to any $b \in B$ as she sees fit. Further, $d$ is always active, it cannot be deactivated, redefined, or deleted. Finally, $P$ is the set of processed images the user has provided feedback on. At all times, $P = B \cup d$.

The main challenge of supporting analytic categorization is its flexibility. There are no constraints, predefined rules, or prior knowledge concerning images the user can add to the buckets. Yet the model must ``read the user's mind'', supplying suggestions that are relevant to $B$ in its current state. Moreover, it must do so in interactive time, placing challenging constraints on computational efficiency. However, the payoff for the flexibility is significant: buckets can be fit by the user to a variety of tasks, including T1 -- T3 from the introduction.

II-20's interface-model scheme is depicted in \autoref{fig:ii20_scheme}. The model suggests images that cover the entire exploration-search axis: from pure search through dynamic interactive learning to exploration candidates that take the user to previously unseen parts of the collection. The suggestions come with a \emph{bucket confidence} score that expresses the model's confidence an image belongs to the bucket. This is visualized in the UI as additional information to enhance the decision making.

User agency is a core design tenet for II-20. The user starts in the familiar grid interface and the model is deliberately a black box: there are no required inputs from the user to control the sliding on the exploration-search axis, the model determines everything from the user's actions and an internal assessment of its own performance. When the user feels comfortable, she can switch back and forth between grid and Tetris and engage the fast-forward interaction.

\subsection{Interface}
\label{sec:interface}

II-20 interface is depicted in \autoref{fig:teaser}. The main view has three components. The \emph{image view} occupies the main portion of the screen and displays images from the collection.
The \emph{bucket banner} below the image view shows the buckets that are active. Finally, the \emph{control panel} on the right side of the screen provides bucket management, image view settings, and the fast-forward button.

\subsubsection{Interaction}
\label{sec:interaction}


II-20’s interaction protocol is based on the standard used in interactive learning: the user labels images for the buckets and discards the non-relevant ones, the model learns from those interactions and provides relevant image suggestions. This approach is especially suited for tasks T1 (the user can easily follow multiple categories of relevance) and T3 (given the classifier is able to capture the nuances important to the user). In addition, II-20 provides a small number of exploration candidates not tied to a bucket to increase coverage of the exploration-search axis.

The grid image view is the static, batch mode showing multiple images at once. Due to the familiarity of the grid, it is II-20's default image view mode. It is integrated into the model rather loosely: it waits for the user's explicit feedback (the user selects the bucket to be labelled and the images to be assigned there) and explicit instructions to show more relevant images. Image suggestions for a bucket appear with a dashed border in the bucket's color with brightness proportional to bucket confidence. The grid is resizable, so the user can choose to see more images or more detail. The user can also preview individual images by right-clicking the thumbnail. This enlarges the image and provides ample space for displaying any associated text and metadata.

\subsubsection{Tetris metaphor}
\label{sec:tetris}

The grid is a familiar, time-tested metaphor, but working with a series of grids for too long might be perceived as tedious. In addition to the static, batch-mode grid, we investigate the topic of dynamic, 1-image-at-a-time (D1I) metaphors which show an image for a limited amount of time, after which they get automatically assigned to the bucket suggested by the model unless the user intervenes. A D1I metaphor complements the grid, providing a possibly welcome change of pace, as well as three key strengths:

\begin{itemize}
\item \emph{Tight integration with the model, with a degree of autonomy} --- a well-trained model simply feeds images into the correct bucket on its own and the user interacts only once in a while to correct wrong suggestions.

\item \emph{Potentially lower number of processed images in total} --- the model learns incrementally and the UI only shows the top relevant image, so the user needs to process fewer images in total to get the same number of relevant images. In other words, a D1I metaphor reaches the same or higher precision and recall growth compared to the grid (we evaluate this claim in the experiments).

\item \emph{Focus on detail} --- with one image shown at a time, the user's attention naturally focuses on details of the image in question, making D1I metaphors a natural candidate for applications where the detail decides the analytic outcome, such as medical imaging or security. This complements the grid, which is better at overviewing the collection and broader-category analytics.

\end{itemize}

II-20's instantiation of a D1I metaphor is the new \emph{Tetris} metaphor (shown in \autoref{fig:teaser}) inspired by the famous game. Tetris operates as follows: images flow from the top one at a time, and descend to one of the buckets in the bucket banner. When an image reaches the bucket, it gets assigned to it, the model processes the assignment, and the next image starts flowing from the top. The user can steer the images between buckets by pressing the left and right arrow keys, pause the flow by hitting spacebar and increase/decrease the descent speed by hitting the down and up arrow, respectively. Speed and pause/play can also be controlled by buttons in the control panel. Finally, hitting the I key opens up an overlay with any associated text and metadata.

The model will mostly suggest images for the buckets that are active. These flow in already positioned above their suggested bucket, connected to the bucket by a line in the bucket's colour. Exploration suggestions appear over the discard pile without a connecting line to any bucket: they tend to be from previously unseen areas of the collection, so whilst providing exploration directions, they are likely incorrect.


\subsubsection{Control panel and bucket banner}
Each bucket has an entry in the top part of the control panel. Bucket deactivation is useful whenever the user wants to focus on something else and return to the bucket in the future. A bucket can be activated and deactivated at any time by clicking its name or icon. Deactivating will remove the bucket from the bucket banner, gray it out in the control panel, and pause model suggestions for that bucket. However, the bucket will be preserved and the user can reactivate it again. 

The eye button next to the bucket name in the control panel opens up the bucket view, showing all bucket images in a grid. The grid can be switched between 3 (default) and 1 images per row, toggling between more images and more detail. The brightness of an image border is proportional to bucket confidence. The bucket view allows sorting by bucket confidence and the time the image was added to the bucket (newest/oldest first). Finally, the bucket view allows transfer of images between buckets with two modes: move and copy. This implements bucket splitting (the user can transfer images in bulk to a new bucket), and bucket redefinition (moving images between bucket triggers model retraining on both the sending and the receiving bucket), both crucial for analytic categorization flexibility. The edit button allows renaming the bucket. The trash bin button deletes the bucket.

The bucket banner provides a quick overview of the state of the buckets. It shows the number of images in the bucket, as well as bucket archetypes, i.e., the images that the model thinks best represent the bucket. Their number is determined by the screen space available (at least one will be shown for each bucket). The user can thus quickly gauge if the model understands her bucket definition.

\subsection{Machine model}
\label{sec:model}

II-20's model's pipeline for suggesting relevant images is depicted in \autoref{fig:sugg_diag}. The core (employing just the black-coloured steps) is simply the interactive learning pipeline. II-20's model enhances it significantly, producing exploration and search suggestions dynamically by monitoring its own performance without direct involvement of the user.

\subsubsection{Data structures}
To extract image features, we use the ImageNet Shuffle deep neural net \cite{imagenet_shuffle} with 4437 concepts representing visual presence of nouns in the image. We extract two feature representations: the concept representation with the 4437 concepts, i.e., recording the net output, and the abstract representation with the output of the second fully-connected layer containing 1024 dense, abstract features that encode the same semantic information (meaningless to the user, but suitable for indexing). The features are used to construct two key data structures.

Firstly, the \emph{collection index}, establishing an efficient semantic similarity structure on the collection. To compute the index, we employ product quantization (PQ) \cite{pq} on the abstract representation. PQ splits the feature matrix column-wise into $m$ equally-sized submatrices (in our case, $m=32$, i.e., 32 submatrices of 32 features each), then quantizes each submatrix using $k$-means (in our case, $k=\min(1024, \sqrt{n})$, where $n$ is the number of images in the collection), preserving the centroid coordinates for each subquantizer. The PQ representation of each image is the concatenation of subquantizer centroid IDs, with lookup tables of centroid distances set up for quick similarity search.

Secondly, the \emph{interactive learning representation} built using Blackthorn \cite{blackthorn}. We use the concept feature representation, compressed using Blackthorn to preserve the top 25 concepts by value per image. This number is deliberately larger than the recommendation of 6 \cite{blackthorn} due to our concept dictionary being $\sim$4x the size of theirs. The resulting sparse representation preserves image semantics and reduces the size by more than 99 percent. II-20's prototype uses \texttt{scikit-learn} \cite{scikit-learn} which directly supports efficient sparse matrix computations.

\subsubsection{Model components}
\label{sec:sugg_comps}

\begin{figure}
    \centering
    \includegraphics[width=\columnwidth]{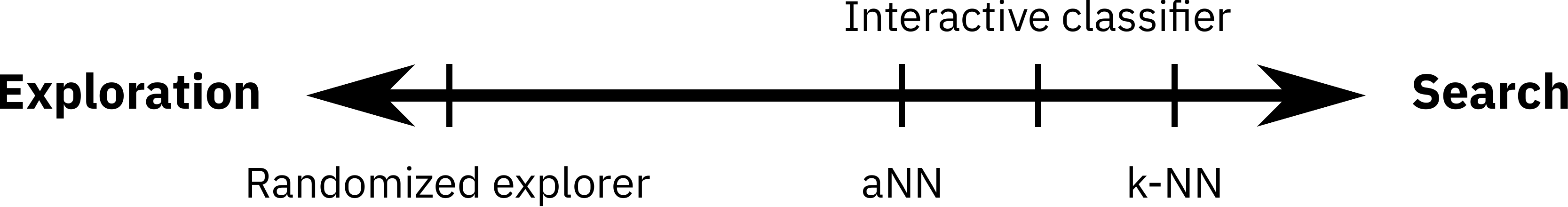}
    \caption{The position of II-20 model components on the exploration-search axis.}
    \label{fig:model_es}
\end{figure}

To cover the entirety of the exploration-search axis, II-20's model has three components capable of suggesting images (as outlined in \autoref{fig:ii20_scheme}): the interactive classifier, nearest-neighbour search, and the randomized explorer. The position of each component on the exploration-search axis is shown in \autoref{fig:model_es}.

The \emph{interactive classifier} maintains a linear SVM model $\sigma_b$ for each non-empty bucket $b \in B$. This classifier choice is consistent with the state of the art in interactive multimodal learning \cite{blackthorn, exquisitor}, linear SVM exhibits good performance in interactive time on even very large datasets. The interactive classifier's suggestions are the top images $i \in I$ by classifier score ($score(\sigma_b, i)$). Since each interactive classifier is explicitly tied to a bucket, it is also used to compute bucket confidence, the belief that an image $i \in I$ belongs to bucket $b \in B$:

\begin{equation}
    conf(i, b) = \min(\max(\frac{score(\sigma_b, i)}{\max_{i_b \in b}(score(\sigma_b, i_b))}, 0), 1)
\end{equation}

As described in \autoref{sec:interface}, bucket confidence is used throughout II-20's UI to provide additional information about the model's reasoning. Easy translation to bucket colour brightness is the reason bucket confidence is confined to the $[0, 1]$ range. If $\sigma_b = \varnothing$, bucket confidence is undefined (but in that case, the model is not suggesting for $b$ anyway).

\emph{Nearest neighbour search} uses the collection index to search for images with the lowest distance to a bucket. It has two modes: $k$-NN ($k$ nearest neighbours) and aNN (approximate nearest neighbours). As shown in \autoref{fig:model_es}, they occupy different positions on the exploration-search axis, and also on the ``exactness vs. computational efficiency'' tradeoff: the $k$-NN mode is more exact, but requires a full $k$-NN matrix of the dataset (especially difficult for datasets of $>$1M images), whereas the aNN mode is more randomized, but utilizes no precomputed structures. We experimentally evaluate both modes.

The $k$-NN mode relies on a precomputed $k$-NN matrix that records 10 nearest neighbours by PQ distance for each image in the collection. To produce suggestions for bucket $b$, the $k$-NN mode uniformly samples images from the set of all recorded neighbours of the images in $b$ (for computational efficiency reasons, if $|b| > 50$, the neighbours of a uniform random sample of size 50 drawn from $b$ are used instead).

The aNN mode first uniformly samples 50000 candidates from all unseen images. Then, it computes their distance to up to 25 images in $b$ (sampling uniformly if $|b|$ > 25). The distance of each candidate to the bucket is the minimal distance between itself and any image in the bucket (sample). Finally, it returns the top candidates sorted by the distance to bucket $b$ in ascending order. The aNN sample caps of 50000 and 25 were chosen to preserve interactive response time.

Finally, II-20 has a \emph{randomized explorer} component to support exploration. It suggests random images that are as far away from what the user has already processed as possible. This allows quick semantic traversal to the unseen parts of the collection. The randomized explorer first randomly samples candidate suggestions from all unseen images. Then, it sorts the candidates by maximum distance to $P$: the distance of each candidate to $P$ is equal to the minimal distance to any image in $P$. The top images in the sorted set are the randomized explorer suggestions. The number of candidates is a performance-bounded parameter: the larger without violating interactive response time, the better. In II-20, it is set to 100 times the number of requested suggestions. 

\subsubsection{Bucket model}
\label{sec:bucket_modelling}

To model buckets, II-20 maintains three extra sets of images per bucket. Firstly, bucket suggestions ($S_b$), i.e., all images suggested for bucket $b \in B$ throughout the session. Secondly, correct bucket suggestions ($C_b)$, all images in $S_b$ which were also subsequently added to the bucket by the user. Thirdly, wrong bucket suggestions ($W_b$), all images in $S_b$ which were then discarded or added to a different bucket. Further, $S_b^{class}$ and $S_b^{nn}$ denote the suggestions produced by the interactive classifier and nearest neighbour search, respectively (similarly for $C_b$ and $W_b$). Let $\llbracket\cdot\rrbracket_{w}$ denote the sliding window operator, which selects exactly those images added in the last $w$ interaction rounds to an image set.

\subsubsection{Suggesting relevant images}
\label{sec:suggs}

\begin{figure*}
    \centering
    \includegraphics[width=0.85\textwidth]{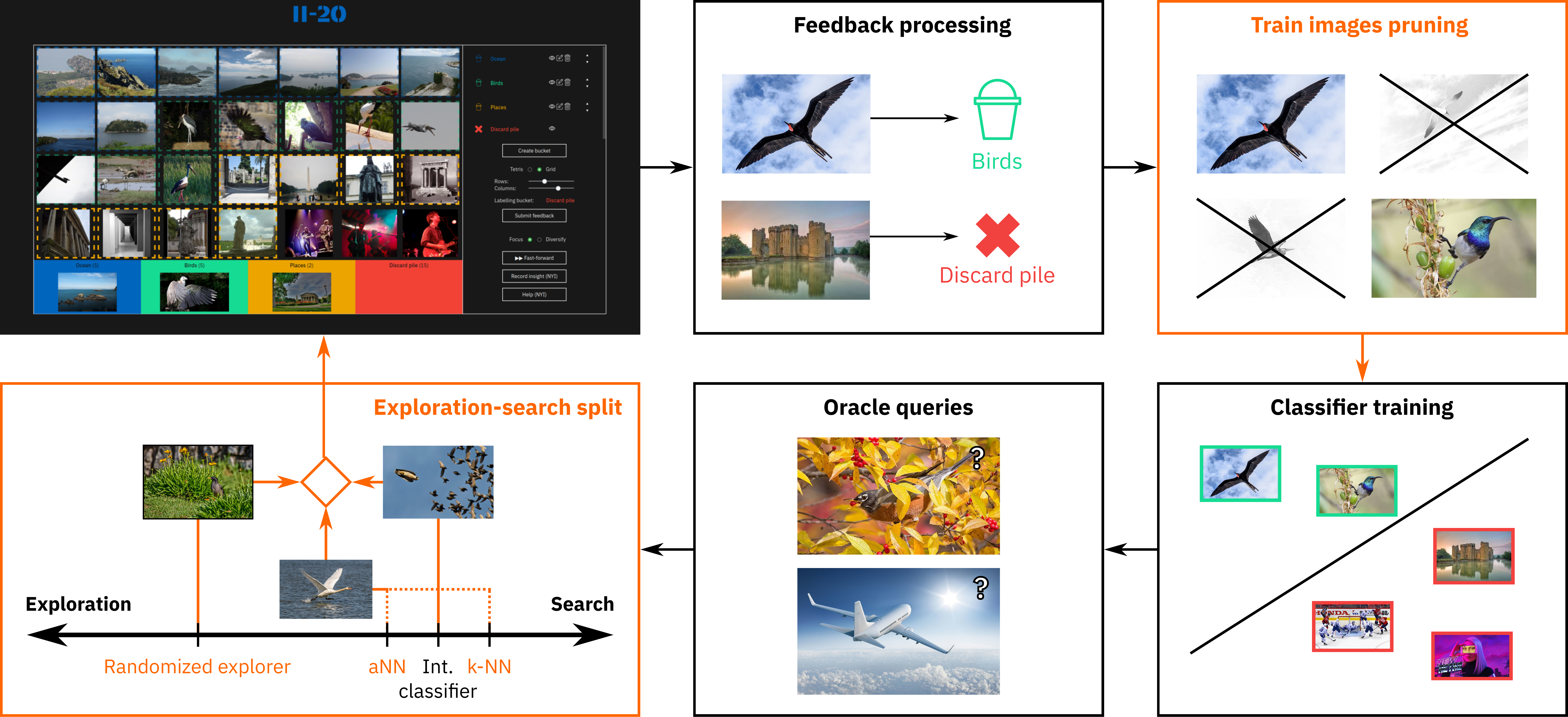}
    \caption{The procedure for suggesting relevant images: feedback is collected from the interface, processed, a new interactive classifier is trained, and then suggestions covering the entire exploration-search axis are produced. The components innovated by II-20 are coloured orange.}
    \label{fig:sugg_diag}
\end{figure*}

The relevant images suggestion procedure takes two inputs: Firstly, user feedback ($F$), a set of key-value pairs with an image as key and its user-assigned bucket as value. Secondly, $s_b$, the number of requested suggestions for each bucket. The suggestion procedure (see \autoref{fig:sugg_diag}) for each bucket operates as follows.

\textbf{Feedback processing} --- Establish $F_b$, the set of all images concerning bucket $b$ in $F$. Split the feedback into positive (images suggested for bucket $b$ and added there by the user), neutral (images added to bucket $b$, but not suggested for it), and negative (images suggested for $b$, but added elsewhere); process each separately. Positive and neutral feedback images are added to $b$, negatives are added to $W_b$.

\textbf{Train images pruning} --- By default, $\sigma_b$ uses all images in $b$ as positive training examples. For increased quality, it may be worthwhile to prune the training set. Generally, the more data, the better, but reinforcing the importance of archetypal images or clarifying the decision boundary might lead to increased performance. To that end, we propose three strategies to construct the positive training set for $\sigma_b$ (if $\sigma_b = \varnothing$, II-20 falls back to the default of taking all images from $b$):

\begin{itemize}
    \item \emph{Relevance feedback} --- The $n_{tr}$ images from $b$ with the highest score according to the current $\sigma_b$, emphasizes the archetypes.
    \item \emph{Active learning} --- The $n_{tr}$ images from $b$ with the lowest score according to the current $\sigma_b$, focuses on the decision boundary between the bucket and the remainder of the collection.
    \item \emph{Hybrid} --- $\frac{n_{tr}}{2}$ relevance feedback images and $\frac{n_{tr}}{2}$ active learning images are obtained, the result is the union of the two sets. Trims images that are neither archetypal nor near the decision boundary.
\end{itemize}

We experimentally compare all four strategies with various $n_{tr}$ to each other and to the default setting (simply taking all images from $b$).

\medskip

\textbf{Classifier training} --- If $b \neq \varnothing$, train the classifier. The set of positives is taken from the previous step, the set of negative training examples is initialized to $W_b$. For classifier robustness, we want at least twice as many negatives as positives. If that is not the case, the negatives are supplemented with a random sample of the images in the discard pile, and if that is still not enough, they are filled to the desired size by a random sample from all images in the collection.

\textbf{Null classifier case} --- If $\sigma_{b} = \varnothing$, return $s_b$ randomized explorer suggestions.

\textbf{Oracle queries} --- Employing active learning often leads to improved classifier quality whilst reducing the required number of user interactions \cite{settles_al, aggarwal_al}. II-20 must chiefly employ relevance feedback, as the user is looking for relevant images, but it might help to ask the user (= the oracle) for judgment on a a couple of difficult images. An oracle query means that instead of the image with the highest $\sigma_b$ score, II-20 shows an image with the score closest to 0 (= nearest to the decision boundary) and marks it with a question mark. Let $o$ denote the proportion of oracle queries within suggestions ($o=0$: pure relevance feedback, $o=1$: pure active learning). II-20 produces oracle queries by replacing each suggestion with an oracle query with probability $o$. Then, $s_b$ is reduced by the number of oracle queries such that the correct requested number of suggestions is maintained. In the experiments, we vary $o$ to gauge the benefits of employing active learning.

\textbf{Exploration-search split} --- The model decides the proportion between classifier, nearest neighbour, and randomized explorer suggestions based on the precision achieved by the classifier ($p^{class}$) and nearest neighbour search ($p^{nn})$ in the last $w$ interaction rounds ($w$ is a parameter subject to experimentation):
\begin{equation}
    p^{class} = \frac{\llbracket C_b^{class} \rrbracket_w}{\llbracket S_b^{class} \rrbracket_w}
\end{equation}

\begin{equation}
    p^{nn} = \frac{\llbracket C_b^{nn} \rrbracket_w}{\llbracket S_b^{nn} \rrbracket_w}
\end{equation}

If $\llbracket S_b^{class} \rrbracket_w = \varnothing$, $p^{class} := 1$ (similarly to $p^{nn}$). For each suggestion to be produced, roulette selection is performed. A uniform random number $r \in [0, 1)$ determines the suggestion source:
    \begin{itemize}
        \item $0 \leq r < p^{class}$: interactive classifier
        \item $p^{class} \leq r < p^{class} + (1 - p^{class})\cdot p^{nn}$: nearest neighbour search
        \item $p^{class} + (1 - p^{class})\cdot p^{nn} \leq r < 1$: randomized explorer
    \end{itemize}
    
In other words, the percentage of interactive classifier suggestions is equal to current precision. Should the interactive classifier start faltering, nearest neighbour search comes in, shifting the position on the exploration-search axis. If neither provides meaningful suggestions, both $p^{class}$ and $p^{nn}$ fall to zero and most of the suggestions will be produced by the randomized explorer, which is traversing to yet unseen parts of the collection. Over a couple of exploration rounds, new bucket images (or even buckets) will hopefully manifest, the sliding window will ``forget'' the bad streak and the analytics will shift toward search again. Task-wise, this enables the model to support tasks incl. T1 -- T3 and allows the user to shift between them on demand (e.g., structuring the collection at first, and looking for needles in the haystack later).

\textbf{Final suggestions} --- Based on the exploration-search split, image suggestions are produced by each of the model components, concatenated with the oracle queries, and returned to the user.


\subsection{Fast-forward}
\label{sec:ff}

The fast-forward interaction quickly expands a bucket using the current model. Fast-forward takes two inputs from the user: the bucket to be expanded ($b_{ff} \in B \cup d, b_{ff} \neq \varnothing$), and the number of images to be added to $b_{ff}$ (denoted $n_{ff}$). As the shorthand notation, we propose ``fast-forwarding $b_{ff}$ by $n_{ff}$''.

After receiving the input, the model directly adds the top $n_{ff}$ images by interactive classifier score to $b_{ff}$. The user is immediately taken to the bucket view with the fast-forwarded images shown at the top of the grid, marked with the fast-forward symbol (double right-pointing triangle). The user can review the fast-forwarded images and transfer the incorrectly-added ones to the discard pile. Note that the fast-forwarded images have already been added to the bucket: not interacting with them will keep them in the bucket, i.e., fast-forward does not merely provide $n_{ff}$ regular suggestions. Closing the bucket view commits the fast-forward; the images will subsequently appear as regular images.

Fast-forwarding brings the following advantages:
\begin{itemize}
    \item \emph{Good model = sped up session} --- Fast-forward provides a gear shift for the session: it is much faster than producing the same number of regular suggestions, regardless of the metaphor. This is useful whenever the user wants to quickly focus on a bucket and expand on insight related to it, without having to grind the broader analytics session to a halt. By enabling this focus shift, fast-forward greatly enhances II-20's capabilities w.r.t. task T2.
    \item \emph{Responsive} --- The model processing part of a fast-forward always completes in interactive time, regardless of $n_{ff}$, due to the model scoring all images in the collection whenever producing suggestions (\autoref{sec:suggs}). The final list of fast-forwards is produced by simply trimming the list to $n_{ff}$, which is computationally trivial.
    \item \emph{Easy discarding} --- Fast-forwarding the discard pile allows the user to quickly dispose of large chunks of non-relevant data, which comes in handy e.g., whenever she has not received relevant suggestions for a while (discards provide valuable negative examples to the model). Model judgments on which images are not relevant tend to be more reliable than on the relevant ones, allowing setting a large $n_{ff}$.
    \item \emph{Semantic} --- ``Fast-forwarding a bucket'' is a comprehensively, clearly defined interaction universally usable across domains of expertise which directly translates to a model adjustment. As such, it answers the call for more semantic interactions \cite{zahalka14, sem_interactions}.
\end{itemize}
\begin{figure*}[ht!]
    \centering
    \includegraphics[width=0.95\textwidth]{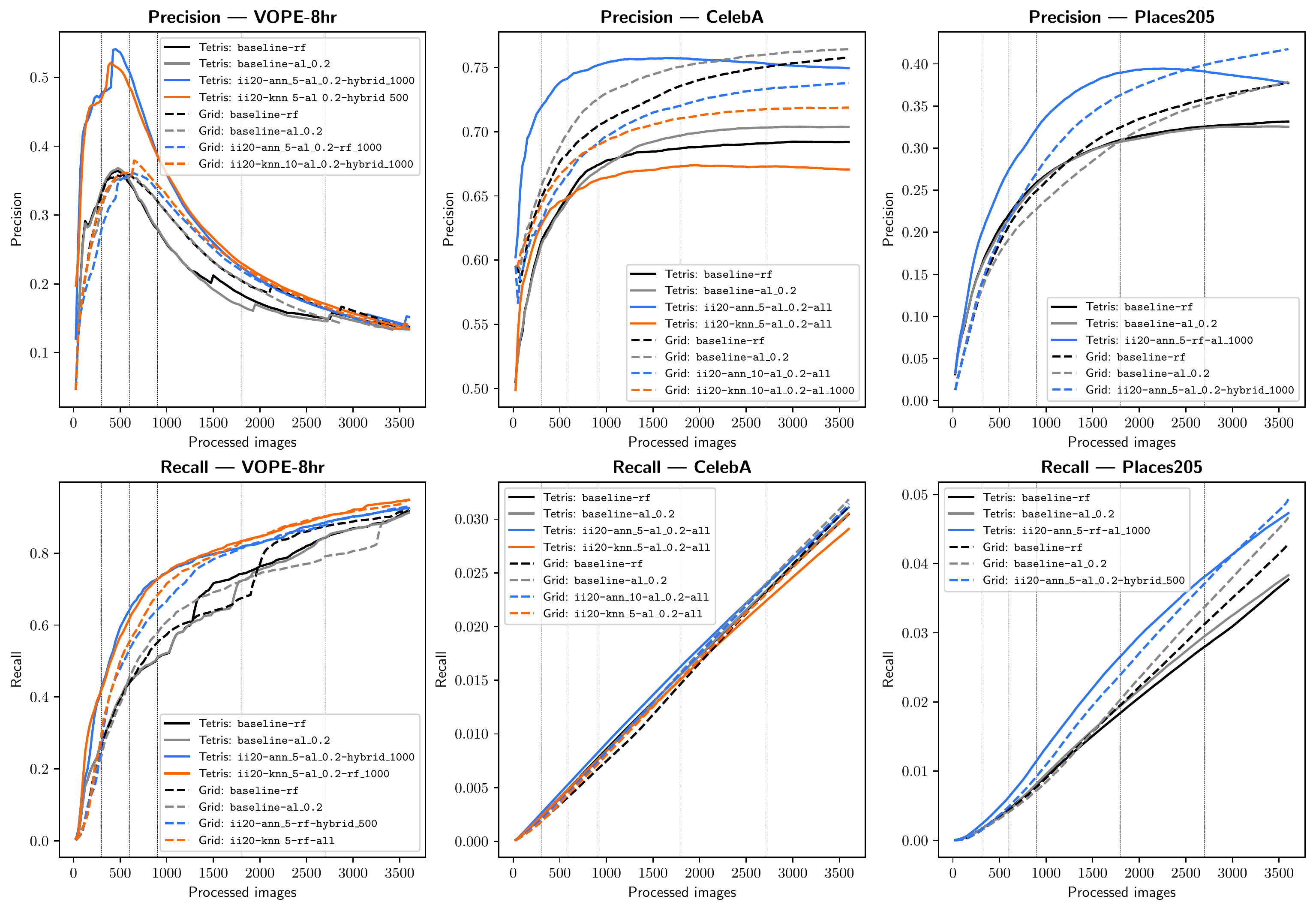}
    \caption{Precision and recall over the number of images processed by the actor, with $x$-ticks at 300, 600, 900, 1800, and 2700 images, corresponding to 5, 10, 15, 30, and 45 minutes at 1 image processed per second.}
    \label{fig:pr}
\end{figure*}

\section{Experimental setup}
\label{sec:exp_setup}

We evaluate II-20 twofold: firstly, we verify the analytic quality of the model with automated experiments, secondly, we perform an open-ended user study gauging II-20's usability and ability to provide insight. In addition, we also report on time complexity.

\subsection{Datasets}
\label{sec:datasets}

We have selected three datasets with different analytic niches: VOPE-8hr, a needles-in-the-haystack dataset with a clear associated real-world task (used for both experiments), and two computer vision benchmark datasets that we use for the automated experiments: CelebA, a portrait dataset with high binary annotation coverage, and Places205, a large-scale scene recognition dataset with categories of varied granularity.

\emph{VOPE-8hr} is a dataset on the topic of violent online political extremism (VOPE). VOPE-8hr was constructed for a real use case: the video analytics component of VOX-Pol \cite{voxpol}, a European network of excellence project connecting social science and forensics research focused on combatting VOPE. The dataset comprises 8 hours of video. 8\% is VOPE content from 3 categories: neo-Nazi, Islamic terrorism, and Scottish ultra-nationalism. 28\% of the content is ``red herring'' content, which exhibits some visual similarity to the VOPE content, but is safe (e.g., comedy skits featuring Nazi paraphernalia in a mocking manner). The rest, 64\% of the content, is fluff, ranging from gaming streams through feature-length films to fashion and football documentaries. We have extracted 1 frame per 3 seconds of video, resulting in a dataset of 9618 images. VOPE-8hr is a challenging dataset: only a small part is relevant (VOPE), and it is obfuscated by three times as much red herring content. The clearly-defined needles-in-the-haystack task and its basis in a real-world use case makes VOPE-8hr very suitable for insight-based evaluation \cite{north_insight, north_insight_eval}.

\emph{CelebA} contains 202K face images annotated with 40 binary attributes such as ``eyeglasses'' or ``wearing hat'' \cite{celeba}. There can (and often are) be more attributes per image, resulting in a large overlap of image sets per attribute. CelebA is a narrow-domain dataset. \emph{Places205} comprises 2.5M scene images, each from one of 205 scene categories \cite{places205}. Places205 brings the challenge of scale (it is not trivial to process 2.5M images interactively), as well as variation in the scope of individual categories: some are quite general (e.g., ``ocean'' or ``office''), some more nuanced (e.g., ``herb garden'' or ``shoe shop'').

\subsection{Automated experiments}
\label{sec:exp_auto}
The automated experiments aim to answer these research questions:
\begin{itemize}
    \item[Q1)] Does II-20's model yield better performance (precision, recall) than classic interactive learning?
    \item[Q2)] How does the Tetris metaphor perform in comparison to the grid?
    \item[Q3)] What parameter configuration of II-20 performs the best?
\end{itemize}

The experimental baseline is Blackthorn \cite{blackthorn}, the state of the art in interactive learning, in two oracle strategy variants. The first, \texttt{baseline-rf}, employs pure relevance feedback ($o = 0$), the second, \texttt{baseline-al\_0.2}, is an active-learning modification with $o=0.2$. Blackthorn is a pure relevance feedback approach, the active learning variant is an adaptation that allows evaluating mixing in active learning.

The baselines are pitted against II-20, varying the parameters defined in \autoref{sec:model} independently. Firstly, the nearest neighbour mode: \texttt{ann} on all three datasets, \texttt{knn} on VOPE-8hr and CelebA (Places205 is too large for $k$-NN matrix computation). Secondly, $w \in \{5, 10\}$, the number of interaction rounds in the exploration-search split. Thirdly, $o \in \{0, 0.2\}$ (\texttt{rf}, \texttt{al\_0.2}), the proportion of oracle queries within the suggested images. Fourthly, the train pruning strategies: \texttt{all} (no pruning), \texttt{rf} (relevance feedback), \texttt{al} (active learning), \texttt{hybrid}. Finally, $n_{tr} \in \{100, 500, 1000\}$, the number of bucket images to be kept when pruning. Henceforth, \texttt{ii20-<nn\_mode>\_$w$-<oracle>-<pruning>-$n_{tr}$} identifier is used for II-20 configurations.

\begin{figure*}
    \centering
    \includegraphics[width=\textwidth]{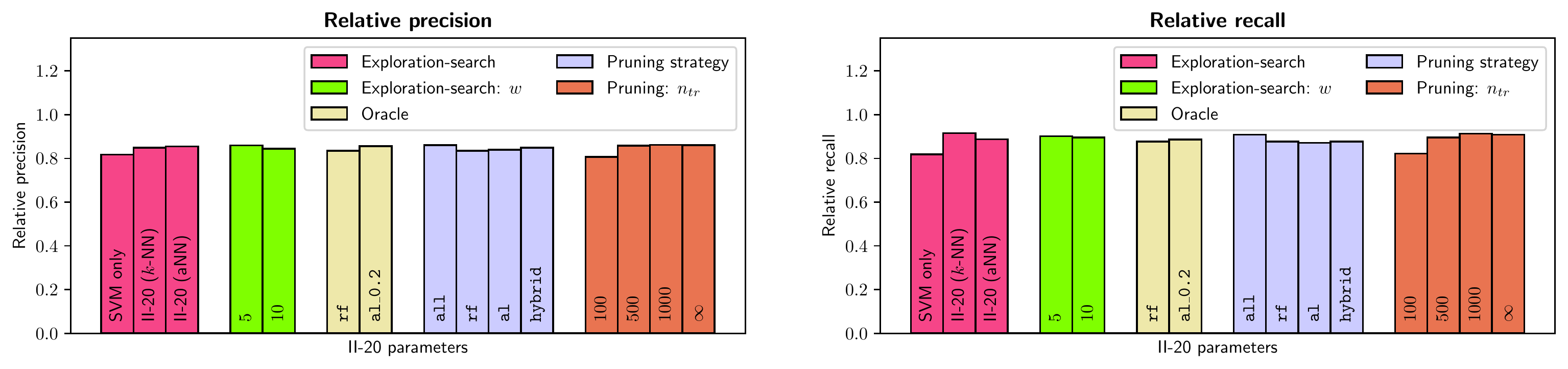}
    \caption{Parameter tuning results (relative precision and recall).}
    \label{fig:par_tuning}
\end{figure*}

For the automated experiments, we employ an enhanced version of the analytic quality evaluation protocol \cite{aq}: artificial actors interact with II-20 in place of a user, putting relevant images in buckets and discarding the non-relevant ones, and we report their achieved precision and recall over time. These actors base their judgment on ground truth annotations that come with each dataset: VOPE categories in VOPE-8hr, facial attributes in CelebA, and scene categories in Places205. For evaluation purposes, the ground truth is known only to the actors and withheld from II-20 in all sessions; II-20 only sees unannotated images. Each actor considers images from a subset of ground truth annotations as relevant and discards all others. Each annotation is treated as a separate bucket. For the VOPE-8hr dataset, we run the experiment on all combinations, i.e., 7 notions of relevance in total. For both CelebA and Places205, we randomly sample 5 notions of relevance for each bucket cardinality from 1 to 7 (matching the active buckets limit as described in \autoref{sec:ii20}), i.e., 35 notions of relevance for each dataset.

In addition to a notion of relevance, each actor has an inherent error rate $err_{a} \in \{0, 0.2\}$: users can make mistakes in their interactions and it is important to test the robustness of the model. Introducing actor errors not only acknowledges the fact that human users are fallible, but also tests resilience against fast-forward errors overlooked by the user. For each label to be produced by the actor, we sample a uniform random number $r \in [0, 1)$. If $r < err_a$, the actor makes one of the following mistakes (with uniform probability): ignores the image (provides no feedback at all), flips relevance (a non-relevant image will be assigned to a random bucket, a relevant image will be discarded), or confuses buckets if applicable (adds the image to a different bucket than it belongs to). The actors vary $err_{a}$ and the notions of relevance independently, resulting in 14 actors total for VOPE-8hr and 70 actors for CelebA and Places205 each.

The actors interact with II-20 in a flow identical to a real user, i.e., a series of interaction rounds: II-20 presents the actor with images, the actor submits its judgment, II-20 updates its model and starts a new interaction round. This allows automatic simulation of metaphor performance: in Tetris mode, the actor reacts to 1 image at a time, in grid mode, the actor reacts to 25 images at a time (a 5x5 grid). Since the model gets updated between user (actor) judgments, labelling a grid results in a different model --- and different image suggestions --- than running Tetris on 25 images. Note that the actors cannot be matched 1:1 to real user behaviour: they only have a static notion of relevance, they do not truly reason, they simply try to find all relevant images based on ground truth. Still, given that they do exhibit key artifacts of real user behaviour, they are useful for automatic approximation of analytic quality of a large combination of II-20 model's parameters, which would be infeasible to evaluate rigorously in user studies.

\subsection{User study}
\label{sec:exp_us}

The user study aims to answer the following research questions:
\begin{itemize}
    \item [Q4)] What are the main strengths and weaknesses of II-20?
    \item [Q5)] How does the Tetris metaphor fare in the eyes of the participants?
    \item [Q6)] How efficient/useful is the fast-forward interaction?
\end{itemize}

We employ an open-ended insight-based evaluation protocol, in which the users think aloud, recording their insights as they progress with their evaluated task \cite{north_insight, north_insight_eval}. The evaluated system's analytic efficiency is then gauged by analyzing these insights. The II-20 user studies are designed to be remote, so the ``thinking aloud'' is replaced by the users hitting the "Record insight" button and recording whatever is on their mind at any point in their session.

The user study scenario has four steps:
\begin{enumerate}
    \item \emph{Introduction} --- The user is greeted by a description of II-20, analytic categorization, and an outline of the user study. Further, the user is linked to a YouTube tutorial and informed that they can refer back to the video at any time during the user study. The importance of using the ``Record insight'' functionality liberally throughout the session is heavily stressed. No special attention is drawn to the novel functionality (innovated model, Tetris, fast-forward) anywhere for the sake of unbiased feedback.
    \item \emph{Warm-up} --- The user tries II-20 out on a toy dataset (same as in the tutorial video). Her objective is to familiarize herself with II-20 controls. The user is not being recorded in this step.
    \item \emph{User study task} --- After the warm-up, the user performs the evaluated task proper (described in detail below).
    \item \emph{Final questionnaire} --- The user answers 3 open-ended questions: strengths of II-20, weaknesses, and any other comments.
\end{enumerate}

The user study task is conducted on VOPE-8hr, and it closely matches its real use case. The user investigates extremists that post propaganda on the Internet. She has just received data from a suspect's computer and is asked to establish whether they contain VOPE content and if so, what kind. In addition, the user is asked to record insights about any content encountered in order to profile the suspect. The user should (among other insights) be able to establish that there indeed is extremist content. The user is instructed to take as much time as needed and can stop the analytic session at any time. In addition to explicit insight, we record user actions and all II-20's image suggestions.

11 users participated in the user study: 9 computer scientists and 2 robotics experts. 9 participants have a master's degree or higher, 2 participants are master students. None of the users are VOPE domain experts; the role of a digital forensics investigator was a role-playing task for them. None of the users have seen the dataset before.
\section{Results and discussion}
\label{sec:discussion}

\autoref{fig:pr} reports the precision and recall of the evaluated algorithms, split by dataset and metaphor. Each curve is the average across all actors running on the dataset. In each plot, for each metaphor, we report the results of the baselines and the best-performing II-20 variants. The $x$ axis is the number of images processed by the user, which has a direct mapping to time (e.g., considering a hypothetical fast user that processes 1 image per second, the $x$ axis is time in seconds).

II-20 outperforms the baselines on all datasets with respect to both precision and recall, except for the later stages of analytics  on CelebA, where the baselines pull ahead slightly (even then, II-20's performance remains quite acceptable). This makes sense: CelebA has high coverage of the annotations used to construct the actors. There are plenty of positive examples, which increases the reliability of the vanilla interactive learning approach. VOPE-8hr and Places205, however, have more of a needles-in-the-haystack nature: positives are a rarer asset. II-20 is strictly dominant on these datasets, often by a large margin. Therefore, we answer Q1 positively, in favour of II-20.

Comparing the metaphor usage simulations reveals that Tetris indeed shows analytic promise: Tetris outperforms the grid on both precision and recall for most of the session on all three datasets. This validates the ``potentially lower total number of processed images'' strength claimed in \autoref{sec:tetris}. It appears that Tetris's performance is tied to whether II-20's model was used: Tetris works well with II-20, baselines are better off with the grid. Also, there seems to be a breakpoint where switching from Tetris to the grid increases performance. We explain this by the difference in availability of training positives. At first, they are rare, and fine-grained feedback after each image (Tetris) is very beneficial to the model. Later on, there are usually enough positives to train the model, but the remaining ones are trickier to find, so it's beneficial to ``fish'' for more by showing a larger portion of the ranking in the grid. Whenever this stage of difficult positives is encountered, it might be worthwhile for the user to switch to the grid. We answer Q2 by remarking that Tetris certainly has strong analytic potential.

To answer Q3, we have performed parameter tuning and report the results in \autoref{fig:par_tuning}: normalized precision at 900 processed images (15 minutes at 1 image/s) and normalized recall at 1800 processed images (30 minutes at 1 image/s), i.e., early precision, late recall. Each bar reports the average normalized precision/recall across all experiments, metaphors, and datasets. Each normalized value is obtained by dividing the absolute value by the maximum achieved on that dataset. This is done to remove differences in absolute performance between datasets.

The differences are not statistically significant: none of the parameters seem to drastically influence the performance (within the evaluated values). However, certain observations can be made. The parameter tuning confirms what \autoref{fig:pr} shows as well: the aNN nearest neighbour mode edges ahead of $k$-NN. This is fortunate: aNN  does not require a $k$-NN matrix. The shorter exploration-search window came ahead, which hints at confirming the importance of the exploration-search sliding being dynamic. Oracle queries seem to improve the model, and pruning should not be done too aggressively (if at all).


\begin{figure}
    \centering
    \includegraphics[width=\columnwidth]{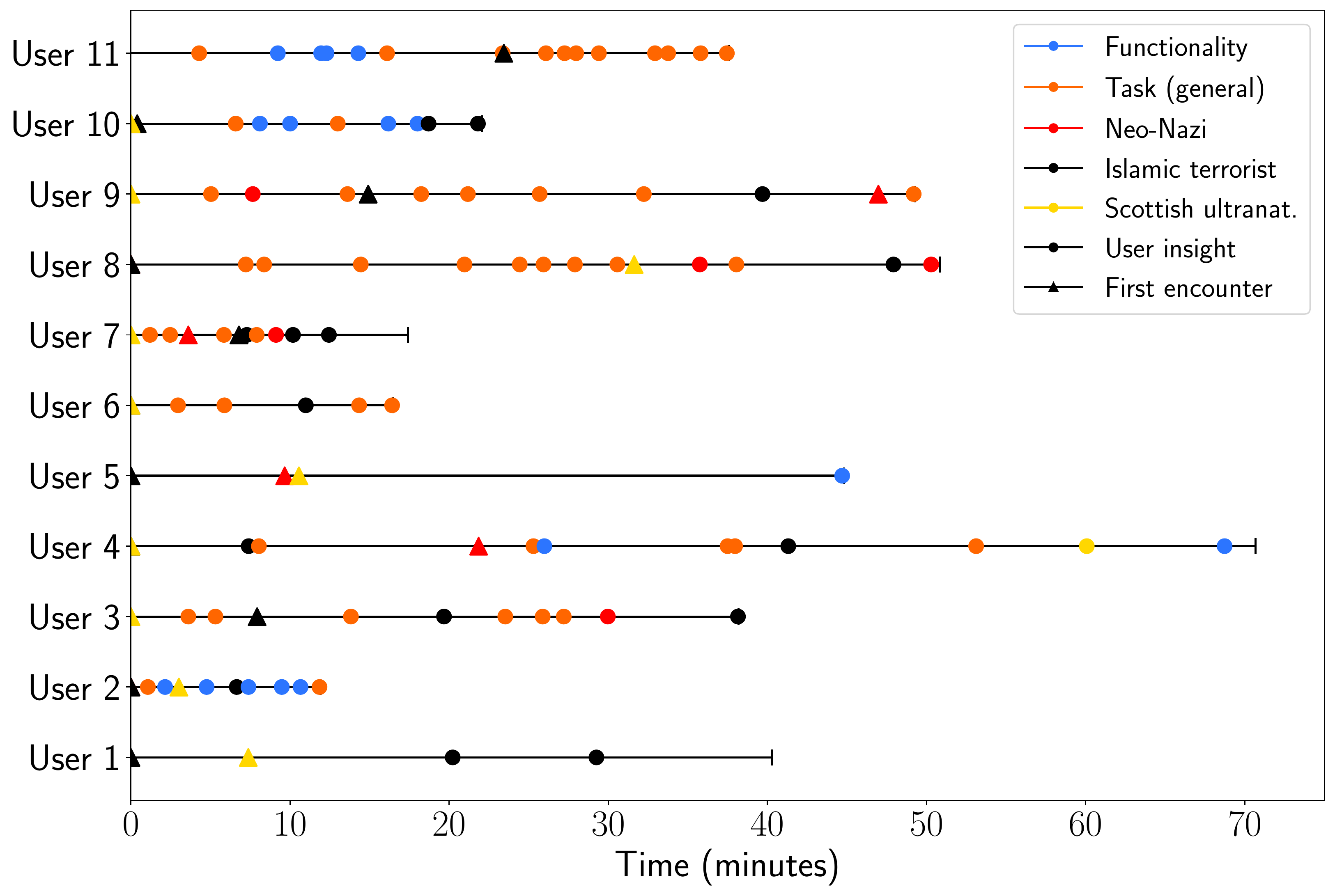}
    \caption{User insights over time by type. Circle markers denote insights, triangular markers the user's first encounter of VOPE content by category.}
    \label{fig:insight}
\end{figure}

\autoref{fig:insight} shows the insights recorded by the user study participants over time, split into five categories. The first are insights related to II-20's functionality, the second are general insights related to the task (for example: ``user plays a lot of video games''), and the remaining three correspond to the VOPE content categories --- neo-Nazi, ISIS, Scottish ultranationalism --- being explicitly referred to by the user (e.g., as one user wrote, ``At the moment I can tell that the suspect does have extremist content from islamic terrorist.''). The triangular markers mark the user's first encounter of an image from a VOPE category.

II-20 was able to sufficiently support the task. All participants were suggested VOPE content, 9 out of 11 participants noted extremism in their insights: all 9 have found Islamic terrorism, 5 have found evidence of neo-Nazi content, and 1 has found out about Scottish ultranationalism (note: this is a difficult, highly contextual category, and none of the users were Scottish). None of the users ended the session prematurely due to being confused or finding the system unusable.

Regarding main strengths and weaknesses, the received feedback is diverse. The strengths reported in the final questionnaire were: intuitive interface and user-friendliness (6 users), full control of the buckets (4 users), and good performance in finding similar images (4 users). The main reported weakness was receiving very similar, non-diverse images from the system (7 users). To an extent, this is an artifact of the dataset (frames extracted from videos, often mono-topical), but that of course does not invalidate the feedback. Other weaknesses include missing progress bar (percentage of content seen) mentioned by 2 users, and diverse feedback by 1 user each, such as the dark design of the UI or unintuitiveness of certain tools (e.g., grid bucket selection and the meaning of discard). Regarding Q4, we conclude that II-20 has succeeded as a tool, it is intuitive and provides good performance, but additional diversification capabilities and UI improvements are needed.

The Tetris metaphor has received a mixed response. It is fair to say that the majority of the users' time was spent in the grid, with 5 of the 11 users swapping to Tetris at any point in the session. That makes sense, the grid is a familiar, useful, and also the default metaphor. Moreover, we have deliberately not drawn any attention to Tetris so as to obtain unbiased feedback. One user has strongly liked Tetris, one has found it not useful, and one stated that it would be much better if it had a static variant with immediate accept/reject. To answer Q5, Tetris comes out as a niche metaphor that polarizes users somewhat: extended functionality and/or implementation of another D1I metaphor (e.g., images flashing on the screen for a certain, user-controlled amount of time that get assigned to the suggested bucket unless the user intervenes) might be beneficial, especially given Tetris's strong analytic performance.

Fast-forward has been used by 8 users at least once. The fast-forwards were by 10--25 images, all concerned user buckets (no user fast-forwarded the discard pile). One user has lauded fast-forward as one of the main strengths of II-20, there has been no negative feedback or suggestions for improvement. Therefore, to answer Q6, we conclude that fast-forward has been shown as a useful, intuitive interaction to those users that have chosen to use it.

\begin{table}
    \centering
    \begin{tabular}{@{}llll@{}}
        \toprule
         & \texttt{baseline} & \texttt{ii20\_ann} & \texttt{ii20\_knn} \\
         \midrule
         \emph{VOPE-8hr} & 0.01 $\pm$ 0.01 s & 0.06 $\pm$ 0.01 s & 0.02 $\pm$ 0.01 s \\ 
         \emph{CelebA} & 0.07 $\pm$ 0.02 s & 1.45 $\pm$ 0.23 s & 0.08 $\pm$ 0.02 s \\
         \emph{Places205} & 0.8 $\pm$ 0.12 s & 2.44 $\pm$ 0.42 s & N/A \\
         \bottomrule
    \end{tabular}
    \caption{Average time per interaction round.}
    \label{tab:time}
\end{table}

Finally, a word on time complexity. \autoref{tab:time} reports the average time per interaction round\footnote{Does not include UI image loading, but that is negligible in II-20: it only loads images with the suggested IDs, which is a very fast DB query.}. Following visual analytics theory \cite{interaction_defined}, II-20's model operates in the direct manipulation ($\leq$ 2--3 s/int. round) regime on all datasets incl. the large Places205 with 2.5M images, on the VOPE-8hr and CelebA datasets even reaching real-time performance ($<0.1$ s/int. round). Using just vanilla interactive learning is unsurprisingly the fastest option, as it is a lower bound: all II-20 configurations perform at least interactive learning every interaction round. Regarding nearest neighbours, $k$-NN is faster, as it frontloads a lot of the computation, but is intractable on large datasets and might yield lower precision/recall as reported above. The aNN mode, while slower, is still interactive even on large data and yields good performance. In the user study, none of the users complained that the system is slow or unresponsive: on the contrary, all received feedback on speed was positive.
\section{Conclusion}
\label{sec:conclusion}
In this paper, we have presented II-20, an approach for multimedia analytics on image collections that contributes towards resolving open challenges in visual and multimedia analytics and closing the pragmatic gap. II-20's new machine model is the first to fully support dynamic sliding on the exploration-search axis without explicit input from the user, and in the automated experiments, it has proven superior to state-of-the-art interactive learning. The Tetris metaphor is a dynamic metaphor with high synergy with the model: an accurate model can ``play the game'' fully autonomously. Tetris has been shown to offer strong analytic numbers and potential. Functionality enhancements, additional metaphor variants, and user study evaluation of the metaphor's impact on the user reasoning capacity would help to establish Tetris and the broader dynamic, 1-image-at-a-time family of metaphors further. The fast-forward interaction expands the family of semantic interactions. It provides an intuitive, fully controllable way to speed up the analytics process.

We especially value that II-20's contributions, in addition to passing the evaluation, have turned out to be considered intuitive by the users. User agency is one of II-20's key design paradigms: the basis is a familiar interface backed by a powerful model, and it's completely up to the user when she wants to engage with the new interactions and interface elements. The II-20 prototype is a complete system, which is available under an open source license for the research community and applied domains alike\footnote{\url{https://github.com/JanZahalka/ii20}}.

II-20's model, even though tested on image data only, is extensible to the multimodal setting (metadata and/or text associated with the images). Features would be extracted for each modality separately. When asked for suggestions, each model component could either split the suggestions between modalities or fuse rankings per modality by rank aggregation (late fusion). As evidenced by recent work on interactive learning \cite{blackthorn, exquisitor}, late fusion does not break the interactive response time requirement even on large datasets. Both approaches are plausible as direct extensions to II-20.

We hope that II-20 has contributed to kicking off truly intelligent and pragmatic multimedia analytics on image collections fit for the new decade.

\acknowledgments{
We would like to thank Paul van der Corput for his comments and suggestions, as well as the test users for their time and feedback. This work was supported by the European Regional Development Fund (project Robotics for Industry 4.0, CZ.02.1.01/0.0/0.0/15 003/0000470).}

\bibliographystyle{abbrv-doi}

\bibliography{ii20}
\end{document}